\documentclass{elsart5p}
\usepackage{graphicx}
\usepackage{amssymb}

\begin{document}

\begin{frontmatter}

\title{Growth and Characterization of Bi$_{2+x}$Sr$_{2-x}$CuO$_{6+\delta}$ Single Crystals}

\author{Huiqian Luo, Lei Fang, Gang Mu, Hai-Hu Wen$^{*}$}

\address{
National Laboratory for Superconductivity, Institute of Physics and\\
National Laboratory for Condensed Matter Physics, P. O. Box 603
Beijing, 100080, P. R. China }

\begin{abstract}

High-quality Bi$_{2+x}$Sr$_{2-x}$CuO$_{6+\delta}$(0$<$x$\leq$0.5)
single crystals have been grown successfully using the
travelling-solvent floating-zone(TSFZ) technique.  The samples
with $x >$ 0.05 are in the underdoped level. The structure of
these crystals was investigated by X-ray diffraction. The
evolution of c-axis lattice parameters with varied \emph{x} is
displayed, which is strongly associated with the behavior of
$T_{c}$. The crystals exhibit superconducting transitions with
$T_c=9$ to $0.8$ K for the samples with $x=0.05$ to $0.20$, while
for samples with $x=0.25$ and above no superconductivity was
discovered down to 1.6 K. The resistivity of samples with
$x$=0.31, 0.40 and 0.50 exhibits a drastic divergence as
temperature approaches 0 K.
\end{abstract}


\end{frontmatter}

\section{Introduction}
Since the high temperature superconductors(HTSCs) were discovered,
there has been no consensus yet on the mechanism of the high $T_c$
superconductivity in these materials. Because of the complex
structure and anomalous behavior in both superconducting and
normal states, it is tempting to conclude that the Landau-Fermi
liquid and BCS theories seem inapplicable to describe the physics
in the normal and superconducting states. The single-layer
copper-oxide superconductor
Bi$_{2+x}$Sr$_{2-x}$CuO$_{6+\delta}$(BSCO) is an ideal material
for the research of mechanism of HTSCs.  This system has a rather
low $T_c \leq 10$ K and a structure with Cu-O conducting layer and
carrier reservoir which is similar to other cuprates
\cite{01,02,03,04,05,06,07,08,09,10}. The advantage to conduct
investigation on this system is that the upper critical field is
rather low (about 25 T) and the superconductivity can be easily
suppressed by applying a magnetic field \cite{11,12,13}, so both
of the superconductivity and normal state characteristics can be
investigated. In addition, the nature of the ground state of
pseudogap is still highly debated \cite{14}, this system may
provide a possible way to reach the ground state of the pseudogap
phase\cite{15}. Comparing to its brother system, the La-doped
Bi-2201 namely $Bi_2Sr_{2-x}La_xCuO_{6+\delta}$ (BSLCO) with
identical structure but much higher $T_c$ ( $T_{c_{max}}\approx$
38 K ), the low $T_c$ of the present system was attributed to a
greater out-of-plane disorder effect induced by a larger mismatch
of the ionic radius between Sr$^{2+}$ and Bi$^{3+}$ than Sr$^{2+}$
and La$^{3+}$\cite{16,17,18}, but more experiments on good quality
single crystals are required to clarify this point.

It was reported that the Bi-based Bi-2201 single crystal could be
grown by the self-flux and KCl-solution-melt
method\cite{01,02,03,04,05,06,07}. However, large size crystals
with high homogeneity and less contamination were still hard to
obtain. On the other hand, the optical travelling-solvent
floating-zone(TSFZ) method could overcome these drawbacks. As far
as we know, only Lin \textit{et. al.} have successfully grown the
La-free bulk superconducting Bi$_{2+x}$Sr$_{2-x}$CuO$_{6+\delta}$
crystals from the non-stoichiometric starting composition of
Bi$_{2.2}$Sr$_{1.9}$Cu$_{1.2}$O$_{6+\delta}$ and
Bi$_{2.35}$Sr$_{1.98}$Cu$_{1.0}$O$_{6+\delta}$ by TSFZ method
under oxygen pressure ($\geqslant$ 2 bar) \cite{08,09,10}.  In
this paper, we report the successful growth of large size
Bi$_{2+x}$Sr$_{2-x}$CuO$_{6+\delta}$ single crystals with high
quality by TSFZ method. The $x$ value in the starting composition
is varied from 0 to 0.50.  The crystals cleaved from the ingots
with $x$=0.04 $\sim$ 0.20 exhibit superconductivity, it seems that
the optimal doping point is at the nominal value of $x$=0.05.
Superconductivity has not been observed at $x$=0.25 and above, and
a much stronger low temperature up-turn of resistivity has been
observed as $x$ is beyond 0.25. A strong correlation between
$c$-axis parameter and $T_c$ is found suggesting an influential
effect on superconductivity by the subtle change of structure.

\section{Experiment and Characterization}

Before the crystal growth, a feed rod with high density is
prepared. The polycrystalline material was prepared by an ordinary
solid reaction method. The starting powders of Bi$_{2}$O$_3$(99.99
\%), SrCO$_3$(99.99\%) and CuO(99.5\%) with the same composition
as the target proportion were mixed by hand in a dry agate mortar
for about 4 hours and pressed into cylindrical rods of $\phi 7
\times 85$ mm under hydrostatical pressure at $\thicksim$70 MPa,
then calcined in a vertical molisili furnace at 780 $^{o}$C for 36
hours under 1 atm atmosphere. The rods were crushed into powder to
be ground again. This procedure was repeated for 4 times to ensure
the homogeneity of the polycrystalline powder. At the last time,
the feed rod was sintered at 850 $^{o}$C for more than 36 h in
order to get a higher density. No crucible was used throughout the
whole process, so there was little contamination for the rods. In
order to get higher density feed rod with sufficient oxygen
content, the premelting was performed under oxygen pressure
$P(O_2)=2$ atm, and the moving speed of mirror stage was 25 $\sim$
30 mm/hr depending on $x$ values. After premelting, a homogeneous
feed rod with $\phi$6 mm in diameter and 60$\sim$90 mm in length
was obtained .

Single crystal growth by the TSFZ method was performed at an
optical floating-zone furnace equipped with four ellipsoidal
mirrors which was produced by the \emph{Crystal Systems
Corporation}. A steep temperature gradient was obtained by using
four 300 W halogen lamp as the heating source. The crystal growth
was under an oxygen pressure in an enclosed quartz tube, two
conditions of oxygen pressure $P(O_2$)=2 atm and 6 atm were
applied, and the O$_2$ flowing rate was about 20 $\sim$ 40 cc/min.
The typical growth rate is about 0.50 mm/hr, which sometimes is
varied between 0.40 mm/hr and 0.60 mm/hr in order to obtain a more
stable floating-zone and larger crystals. The rotation rate is
25.0 $\sim$ 28.0 rpm for the upper shaft and 14.0 $\sim$ 17.0 rpm
for the lower shaft in opposite directions.

In order to check the quality of crystals, the as-grown ingots
were cleaved into many pieces of crystals. Then the single
crystals selected under microscope were characterized by various
techniques. The single crystal surface topography was examined by
scanning electron microscopy (SEM, Hitachi S-4200), and the energy
dispersive X-ray (EDX, Oxford-6566, installed in the S-4200
apparatus) analysis was used to determine the composition of the
crystals. While the X-ray diffraction(XRD) of the crystals was
carried out by a \emph{Mac- Science} MXP18A-HF equipment with
$\theta-2\theta$ scan to see the crystalline quality of the
samples. $K_{\alpha}$ radiation of Cu target was used, and the
continuous scanning range of 2$\theta$ is from 5$^{o}$ to
80$^{o}$. The splitting peaks at high degree angle indicated
different wavelength effects between $K_{\alpha1}$ and
$K_{\alpha2}$ radiation. Before crystal growth, XRD for the powder
of polycrystal was also carried out to check the phase purity of
starting materials. The c-axis lattice parameters were calculated
from XRD patterns of the single crystals, the raw data of XRD was
analyzed by \emph{PowderX} software where the zero-shift and
$K_{\alpha2}$-elimination and other factors were taken into
account \cite{dong1,dong2,dong3}. The superconductivity of the
crystals was measured with AC susceptibility and resistivity based
on an \emph{Oxford }cryogenic system Maglab-EXA-12 and a
\emph{Quantum Design} Physical Property Measurement System(PPMS).
An alternating magnetic field $H=0.1$ Oe was applied perpendicular
to the $ab$-plane at a frequency $f$=333 Hz when the AC
susceptibility measurement was undertaken with the
zero-field-cooling(ZFC) method. The transition temperature ($T_c$)
of the crystals was derived from AC susceptibility curve by the
 point where the real part of the susceptibility becomes flat.
 A four-probe Ohmic contact with low resistance ($<10  \Omega$) on $ab$-plane was used
for the resistivity measurement , and the sweeping rate of
temperature is about 2 K/min from 2 K to 300 K.

\section{Results and discussion}

\subsection{Crystal growth}
The crystal growth was performed under oxygen atmosphere enclosed by
a quartz tube.  At the initial stage of the heating, both of the
feed rod and seed rod were set apart by a distance of a few mm, the
lamp power was increased gradually until both ends of the feed rod
and seed rod began to melt. Then the upper feed rod was moved
downwards carefully until it touched the seed rod and formed the
molten zone. The length and width of floating-zone were adjusted
carefully so that its diameter was almost the same as that of the
feed and seed rods(about 6 mm).

In order to get high quality single crystals, a stable
floating-zone with proper volume should be sustained during the
growth. The parameters such as the power of lamp, growth rate,
rotation rate of shafts, pressure and flowing rate of oxygen
should be carefully chosen. The volume of floating-zone could also
be adjusted carefully by the downwards or upwards slow moving of
the upper shaft. The length of floating-zone, namely the distance
between the feed rod and seed rod, was sensitive to the power of
lamps and thus a proper power of lamp was applied. It could be
changed 0.1\%$\sim$0.5\% to get the optimal length of
floating-zone. While the width of floating-zone was sensitive to
the moving speed of mirror stage, namely, the growth rate of
crystals. Sometimes it was tuned ($< \pm $ 0.1mm/hr) in order to
obtain a more stable floating-zone and larger crystals. At the
same time, both of the upper and lower shafts were rotated in
opposite directions, the rotation rates which determine the convex
shape of solid-liquid interface could also be changed at a
suitable range until the floating-zone was stable enough.
Generally, the rotation rate of upper shaft was relatively faster
to mix the liquid homogeneously, but the lower shaft was rotated
at a lower rate to get a stable floating-zone and large size
crystals. Furthermore, the pressure of oxygen could influence on
the melting point of the feed rod and stability of floating-zone.
Comparing the case between $P(O_2)=2$ atm and $P(O_2)$ =6 atm, we
found no distinct effects on crystal structure and $T_c$. However,
for the same composition of starting material, larger flowing rate
of $O_2$ would lead to little larger crystals but make the
floating-zone more unstable, while higher pressure did not lead to
larger thickness along c-axis of the crystals. This is slightly
different from the previous reports\cite{09,10}, and it will be
discussed in the following.

It is noteworthy that all growth parameters depend on starting
compositions with different $x$ values, because the relative
content of Bi$_{2}$O$_{3}$ influences the melting point of feed
rod and the viscosity of the mixed liquid. To our experience, the
properties of the crystal  mostly depended on Bi content in the
starting material and had little relation with oxygen pressure.
For $x>$0.10, large plate-like single crystals could be obtained
easily, but for 0$<x<$0.10, the crystals became smaller and for
$x=0$ there were only tiny needle-like crystals were obtained. In
addition, for the too much higher and lower viscosity of compound
with $x=0.50$ and $x=-0.20$ (where Sr was more sufficient than Bi)
respectively, the floating-zone collapsed frequently and the
crystal growth was interrupted. This is similar to the observation
reported in ref.\cite{09,10}, while Lin \emph{et.al.} only
successfully obtained large single crystals from the
non-stoichiometric starting composition when Bi content was more
than 2.2. On the other hand, most of as-grown single crystals were
in underdoped level. It could be attributed to the deficiency of
oxygen content of the starting material, because we only sinter
the feed rod in air instead of flowing oxygen and premelt the feed
rod under oxygen pressure $P(O_2)$=2 atm. From our result, it
seems not easy to change the oxygen content by changing the oxygen
pressure during growth. For the same composition of starting
material, crystals grown under $P(O_2)$=0.5 atm , $P(O_2)$=2 atm
and $P(O_2)$=6 atm have almost the same features such as $T_c$ and
size. So we suppose the crystals grown under the same condition
have almost the same oxygen contents, and the x value has a simple
linear relationship with the doping level $p$ where $p=a-bx$. One
of the as-grown ingots with x=0.15 and several cleaved as-grown
crystals with x=0.12, 0.15, 0.20 and 0.31 are showed in Fig.1. The
crystals are sizeable and flat in large area which exhibit the
high quality.

\begin{figure}
  \center\includegraphics[width=3.0in]{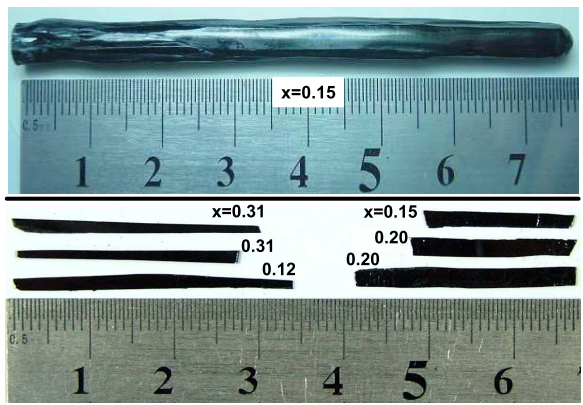}\\
  \caption{(color online)As-grown ingot of BSCO with $x$=0.15 (above)
  and single crystals cleaved from as-grown ingots with x=0.12, 0.15, 0.20 and 0.31 (below).}
  \label{f1}
\end{figure}

\subsection{ Composition and structure of the crystals}

The crystal composition was examined by the energy dispersive
X-ray (EDX) analysis. For each $x$ value, 3$\sim$5 pieces of
as-grown single crystal were selected from the crystals cleaved
from different parts of the as-grown ingot and taken EDX
measurement. One of the typical EDX spectrums for $x$=0.40 is
shown in Fig.2. The crystal composition normalized to Sr=1.60 is
Bi$_{2.41}$Sr$_{1.60}$Cu$_{1.05}$O$_{6.36}$. Note that the
composition of oxygen is not precise as other elements because the
EDX is not sensitive to light atoms. From the result, it can be
seen that the crystal composition is close to the starting
material and Cu has slightly more sufficient content than Cu=1.
For other $x$ values, the results are similar to $x$=0.40 except
for $x$=0.05 and 0.04 because of the inhomogeneity of those
needle-like crystals. However, there is no remarkable difference
between crystals grown under $P(O_2)$=2 and 6 atm. A brief summary
of the properties of the single crystals grown under $P(O_2)$=2
atm is given in Table I.

\begin{figure}
   \center\includegraphics[width=2.4in]{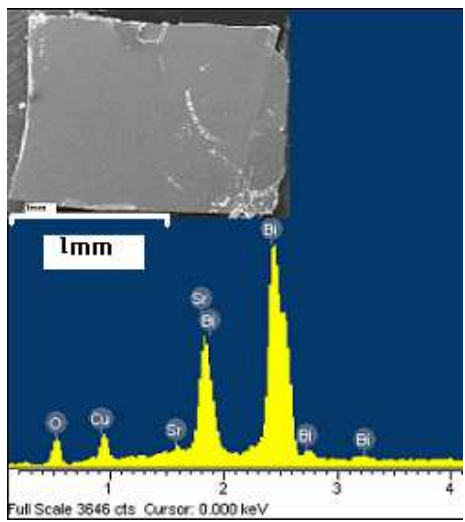}\\
  \caption{(color online)Typical EDX spectrum for the single crystal with $x$=0.40,
  the inset is the SEM photo of this crystal.
 The crystal composition normalized to  Sr=1.60 is
Bi$_{2.41}$Sr$_{1.60}$Cu$_{1.05}$O$_{6.36}$.}
  \label{f2}
\end{figure}

In order to examine the structure of our samples, XRD measurement
was carried out. The typical single crystal XRD patterns are shown
in Fig.3(a). Only sharp even peaks ($00l$) along to $c$-axis could
be observed. The full-width-at-half-maximum(FWHM) of each peak is
about 0.10$^{\circ}$ $\sim$ 0.12$^{\circ}$. The peaks are
splitting at high degree angle due to different wavelength effects
between $K_{\alpha1}$ and $K_{\alpha2}$ radiation.The powder XRD
measurement for crushed BSCO crystals from the ingot was carried
out to check for possible phase decomposition, and XRD for the
polycrystal and starting material after premelting was also
carried out before crystal growth. It is shown in Fig.3(b), all
reflections could be indexed to the tetragonal structure.
Comparing with the XRD patterns for the single crystal, it is
found that the position of peaks from four sets of data coincide
very well. This indicates that the phase composition was not
changed throughout the whole process.

\begin{figure}
  \center\includegraphics[width=3.5in]{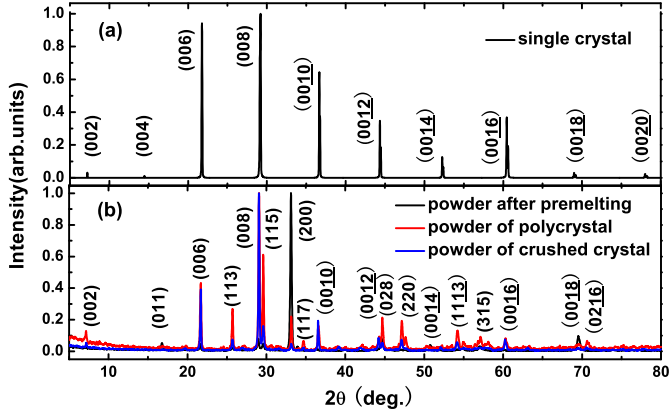}\\
  \caption{(color online)(a)Typical XRD patterns for cleaved single crystal;
  (b)Typical powder XRD patterns for polycrystal, starting material after premelting and crushed crystal. }
  \label{f3}
\end{figure}

The c-axis lattice parameters were calculated from the XRD
patterns of the single crystals, the raw data of XRD was analyzed
by \emph{PowderX} software where the zero-shift and
$K_{\alpha2}$-elimination and other factors were taken into
account \cite{dong1,dong2,dong3}. Some results were shown in Table
I, and the x value dependence of $c$-axis lattice parameters for
the single crystals grown under $P(O_2)$=2 atm was shown in
Fig.6(a). There is a subtle compression of the c-axis parameters
when x increases from 0.20 to 0.31, and the c-axis parameters are
almost constant when x $\leqslant$ 0.20 and x $\geqslant$ 0.31
respectively. This may be attributed to the substitution of
Sr$^{2+}$ (d=1.12 \AA) and Bi$^{3+}$ (d=0.96 \AA), and also may
have a little relation with the influence of monoclinic distortion
and layer modulation. The major factors induced this behavior of
c-axis should be carefully discussed, further experiments on TEM
are proceeding to find them. The $c$-axis is compressed slightly
under the case with $P(O_2)$=6 atm, but the behavior of the c-axis
parameters is similar to that shown in Fig.6(a).

\begin{table}

Table I. A brief summary of the properties of the single crystals
grown under $P(O_2)$=2 atm:
The actual cationic compositions(Bi:Sr:Cu), ratio of Bi/Sr, critical temperature($T_c$), c-axis parameters and crystal-feature.\\
\begin{center}
\begin{tabular}{|c|c|c|c|c|c|}
 \hline \hline
x\qquad   & $Bi:Sr:Cu$   & \quad $Bi/Sr$    &\quad$c(nm)$   &\quad$T_c(K)$  &\quad$Crystal-feature$ \\
  \hline
0.10     & $2.13:1.90:1.25$       & $\quad1.12$      & $\quad2.4656\quad$        & $7.5$  &Needle-like\\
0.15     & $2.27:1.85:1.14$       & $\quad1.23$      & $\quad2.4648\quad$        & $2.4$   &Plate-like\\
0.20     & $2.21:1.80:1.15$       & $\quad1.23$      & $\quad2.4651\quad$        & $0.8$  &Plate-like \\
0.31     & $2.31:1.69:1.17$       & $\quad1.37$      & $\quad2.4530\quad$        & $ns$  &Plate-like\\
0.40     & $2.41:1.60:1.05$       & $\quad1.50$      & $\quad2.4533\quad$        & $ns$ &Plate-like \\
0.50     & $2.36:1.50:1.11$       & $\quad1.57$      & $\quad2.4523\quad$        & $ns$ &Needle-like \\

\hline \hline
\end{tabular}
\end{center}\label{tab:tableI}
\end{table}

\subsection{Superconductivity}

The AC susceptibility was measured on more than 20 pieces of
crystals which were cleaved from different parts of the as-grown
ingot for each $x$ value. Fig.4 shows three typical curves of the
AC susceptibility for the crystals with $x$=0.14, 0.12 and 0.10,
the $T_c$ was defined as the point where the real part of the
susceptibility becomes flat, namely, the onset point. The
superconducting transition width($\Delta T_c$=$T_c$(90\%) -
$T_c$(10\%)) of the crystals is very sharp, about 0.3 $\sim$ 0.8
K. For the needle-like crystals, $\Delta T_c$ is more than 1.0 K.
However, due to the strong anisotropy of bismuth based single
crystals, it is almost impossible to obtain single-domain as-grown
ingot\cite{10,19,20}. Actually, the orientation of $c$-axis was
always perpendicular to the growth direction. The cleaved surface
was always $ab$-plane, so the thickness of cleaved crystals was
 small. However, it was reported that higher pressure of
oxygen lead to thicker and sizeable crystals\cite{09,10}. But by
using 6 atm oxygen pressure during the growth, the size of
crystals was not improved obviously, and the $T_c$ was increased
less than 0.5 K compared with $P(O_2)$=2 atm. Actually, due to the
random growth direction of crystals, the segregation always
exists, this leads to the slightly difference of $T_c$ of
different pieces cleaved from the same ingot. The curves of the AC
susceptibility for all crystals cleaved from the same ingot were
drawn together, and the $T_c$ for each $x$ value was determined by
statistic method, that is, the average value for samples with
close $T_c$ values.

\begin{figure}
  \center\includegraphics[width=3.0in]{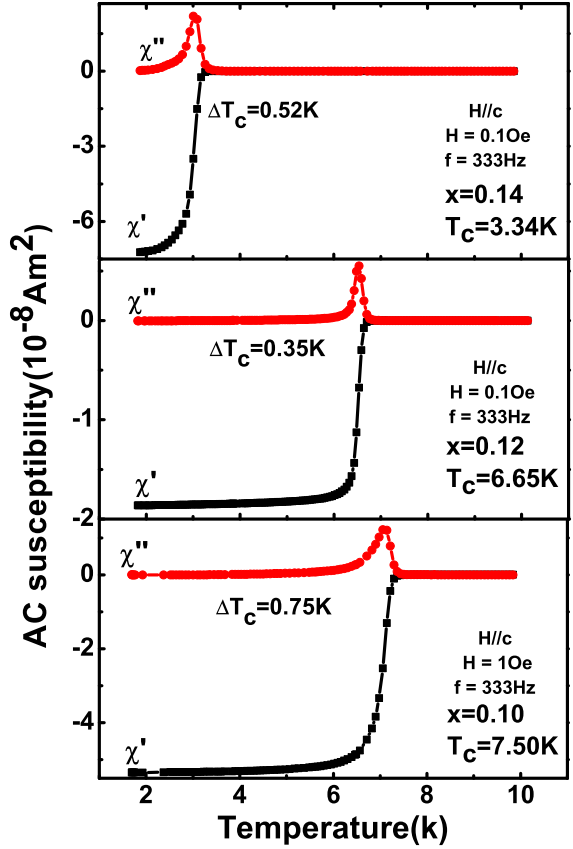}\\
  \caption{(color online)AC susceptibility for $x$=0.14, 0.12, 0.10 single crystals, the $T_c$
was defined as the point where the real part $\chi'$ derates from
the flat part, and $\Delta T_c$=$T_c$(90\%) -
$T_c$(10\%).}\label{f5}
\end{figure}

As known for cuprate superconductors, post-annealing under
different temperature and atmosphere could change the oxygen
content, and thus the doping level as well as $T_c$. But it seems
not easy to change the $T_c$ of Bi-based phase Bi2210 by post
annealing \cite{05,12,23,24}. The crystallinity becomes worse and
phase segretion occurs in the annealing at a high temperature
\cite{24}. It was tried many times to change the oxygen contents
by annealing in flowing oxygen and nitrogen. The annealing
temperature was varied 20 $^{\circ}$C each time from 400 to 600
$^{\circ}$C, and sustained for more than 100 hours each time. But
the curves of the AC susceptibility for the same crystal moved a
little after annealing, the largest change was less than 0.5 K.
Perhaps the proper annealing temperature is more than 600
$^{\circ}$C, but some crystals with big x value melted already
under 600$^{\circ}$C in 1 atm atmosphere. The x value dependence
of $T_c$ was derived from high quality as-grown single crystal for
each $x$ value, which has been shown in Table I and Fig.6(b).

\begin{figure}
  \center\includegraphics[width=3.2in]{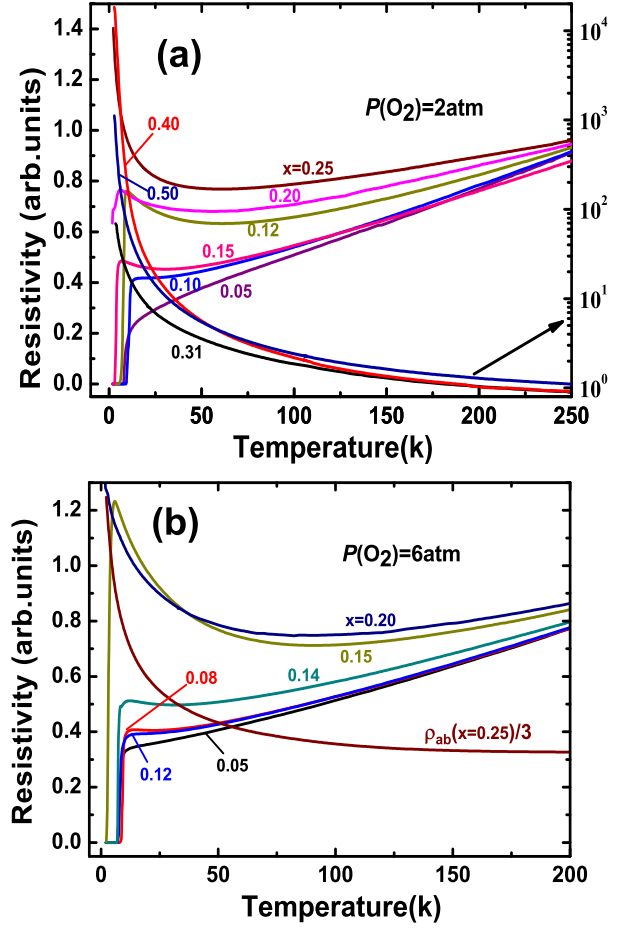}\\
  \caption{(color online)Temperature dependence of resistivity for BSCO single
  crystal. Fig.5 (a) and (b) show the resistivity of BSCO grown under
  oxygen pressure $P(O_2)$=2 atm and $P(O_2)$=6 atm, respectively. Where
  resistivity for the samples with x=0.31, 0.40, 0.50 was
  corresponding to the right axis with a logarithmic scale.
   }\label{f6}
\end{figure}

Fig.5 shows the resistivity of as-grown single crystals, where
Fig. 5(a) and Fig. 5(b) show the resistivity of the single
crystals grown under 2 atm and 6 atm pressure of oxygen,
respectively. All resistivity curves were normalized by
$\rho(T$=280 K) in order to compare with each other. From Fig.5 we
can see that most of the curves show underdoped behavior, except
for $x$=0.05 in Fig. 5(a), which shows overdoped feature. When the
temperature is close to 0 K, the upturn of resistivity becomes
more serious as $x$ value increases. The resistivity for $x$=0.31,
0.40 and 0.50 samples which correspond to the right axis in Fig.
5(a) become drastically divergent when $T$ approaches 0 K.
Moreover, the temperature dependance of the crystal resistivity
under magnetic field up to 9 T was measured recently. A negative
magnetoresistance effect was observed in the weak localized region
of resistivity, which may be explained by the delocalization
effects.\cite{21,22}

 The evolution of $T_c$ with varied $x$ was shown in
Fig.6(b), the $T_c$ values were derived by statistic method as
described above. It is known that the hole doping decreases as
 $x$ value increases, thus the substitution of Sr$^{2+}$ with
Bi$^{3+}$ enhances the electron doping and reduces the hole
concentration. It is obvious that all samples with $x$ varied from
0.05 to 0.50 were in the underdoped region, this can be confirmed
further by resistivity measurement. For the samples with $x$=0.05
which is close to optimal doping level, $T_c$ varies in wider
range in different parts of the same ingot. And for the samples
with $x$=0.04, the situation is similar to $x$=0.05, but most
samples are in the overdoped region. In addition, the $T_c$ of
samples with $x$=0.20 was derived by extrapolating the resistivity
to zero at low temperatures, and for samples with
$x\geqslant$0.25, no superconductivity was observed down to 1.6 K.
Fig. 6(b) shows the relation of $T_c$ vs. $x$ and it seems that a
half dome shape emerges.  While Fig. 6(a) shows the corresponding
behavior of c-axis parameters. It is interesting to note that
there is a strong correlation between $T_c$ and the c-axis
parameters. The vanishing of superconductivity starts from the
point where the contraction of c-axis occurs, which suggests that
the superconductivity is strongly influenced by a subtle change of
the structure.

\begin{figure}
  \center\includegraphics[width=3.0in]{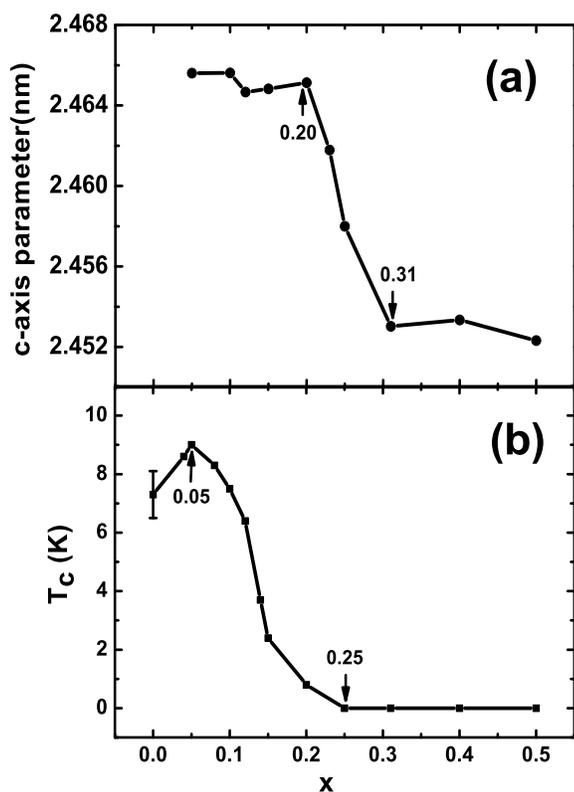}\\
  \caption{The doping dependence of (a) c-axis parameters and (b) $T_c$ of the as-grown BSCO single crystals.
   }\label{f7}
\end{figure}

\section{Conclusions}

High quality underdoped Bi$_{2+x}$Sr$_{2-x}$CuO$_{6+\delta}$
single crystals were successfully grown by TSFZ method. Higher
pressure of oxygen during growth and post-annealing have little
effects on the $c$-axis parameters and $T_c$. The
superconductivity vanishes at $x$=0.25, and the resistivity of
crystals with $x\geqslant$ 0.31 become drastically divergent with
$T$ down to zero. There is a strong correlation between the c-axis
parameters and $T_c$ suggesting a influence on superconductivity
by a subtle change of the structure.

\section{Acknowledgement}
This work is supported by the National Science Foundation of
China, the Ministry of Science and Technology of China (973
project: 2006CB601000, 2006CB0L1002), the Knowledge Innovation
Project of Chinese Academy of Sciences (ITSNEM). The authors thank
the great helps from Prof. H.Chen and Prof. C.Dong for the XRD
measurement and data analysis, and thank the helpful discussion
with Prof. Y.Z.Zhang, Prof. L.Shan and Prof. R.Cong at IOP, CAS.
\\

$^{*}$Email address of the corresponding author:
hhwen@aphy.iphy.ac.cn


\begin{thebibliography}{00}

\bibitem{01}C. Michel, M.Hervieu, M.M.Borel, A.Grandin, F.Deslandes, J.Provost and B.Raveau, Z.Phys. B \textbf{68},421(1987);

\bibitem{02}L. F. Schneemeyer, R. B. van Dover, S. H. Glarum, S. A. Sunshine, R. M. Fleming, B. Batlogg,
T. Siegrist, J. H. Marshall, J. V. Waszczak, L. W. Rupp, Nature
\textbf{332},422(1988);

\bibitem{03}G.Xiao, M.Z.Cieplak and C.L.Chien, Phys.Rev.B  \textbf{68}, 11824 (1988);

\bibitem{04}R. M. Fleming, S.A.Sunshine, L.F.Schneemeyer, R.B.Van Dover,
R.J.Cava, P.M.Marsh, J.V.Waszczak, S.H.Glarum S.M.Zahurak and
F.J.DiSalvo, Physica C, \textbf{173} (1990)37;

\bibitem{05}K.Remschnig, J.M.Tarascon, R.Ramesh and G.W.Hull, Physcica C, \textbf{175} (1991)261;

\bibitem{06}J.I.Gorina, G.A.Kaljushnaia, V.I.Ktitorov, V.P.Martovitsky, V.V.Rodin,
 V.A.Stepanov and S.I.Vedeneev, Solid State Commun. \textbf{91}, 615(1994);

\bibitem{07}T.Niinae, Y.Ikeda, Y.Bando, M.Takano, Y.Kusano, J.Takada, Physcica C, \textbf{313} (1999)29;

\bibitem{08}M. Matsumoto, J. Shirafuji, K. Kitahama, S. Kawai, I. Shigaki and Y. Kawate, Physcica C, \textbf{185-189} (1991)455;

\bibitem{09}C. T. Lin, B. Liang, M. Freiberg, K. Peters and E. Sch$\ddot{o}$nherr
, Physcica C, \textbf{341-348} (2000)541;

\bibitem{10}B. Liang, A. Maljuk and C. T. Lin, Physcica C, \textbf{361} (2001)156;

\bibitem{11}M.S.Osofsky, \emph{et al.}, Phys.Rev.Lett \textbf{71}, 2315(1993);

\bibitem{12}S.I.Vedeneev, A. G. M. Jansen, E. Haanappel, P. Wyder, Phys.Rev.B \textbf{60}, 12467(1999);

\bibitem{13}F.Bouquet, L.Fruchter, I.Sfar, Z.Z.Li and H.Raffy, Cond-mat/0512093(2006);

\bibitem{14} A. Kanigel \emph{et al.}, Nature Physics \textbf{2}, 447(2006)Letters;

\bibitem{15}Hai-Hu Wen, Lei Shan, Xiao-Gang Wen, Yue Wang, Hong Gao, Zhi-Yong Liu,
 Fang Zhou, Jiwu Xiong, and Wenxin Ti, Phys. Rev. B \textbf{72},134507(2005);

\bibitem{16}H. Eisaki, N. Kaneko, D. L. Feng, A. Damascelli, P. K. Mang, K. M. Shen, Z.-X. Shen, and M. Greven,
Phys.Rev.B \textbf{69}, 064512(2004);

\bibitem{17}K. Fujita, T. Noda, K. M. Kojima, H. Eisaki, and S. Uchida, Phys. Rev. Lett. \textbf{95}, 097006(2005);

\bibitem{dong1}C. Dong, J. Appl. Cryst. (1999), \textbf{32}, 838-838;

\bibitem{dong2}C. Dong, H. Chen and F. Wu, J. Appl. Cryst. (1999),\textbf{32}, 168-173;

\bibitem{dong3}C. Dong, F. Wu and H. Chen, J.Appl. Cryst. (1999), \textbf{32}, 850-853;

\bibitem{18}A. Sugimoto, S. Kashiwaya, H. Eisaki, H. Kashiwaya, H. Tsuchiura,
Y. Tanaka, K. Fujita, and S. Uchida, Phys. Rev.B \textbf{74}, 094503(2006);

\bibitem{19}B.Liang and C.T.Lin, J.Crystal Growth, \textbf{267}(2004)510;

\bibitem{20}B.Liang and C.T.Lin, J.Crystal Growth, \textbf{237-239}(2004)756;

\bibitem{21}T.W.Jing, N.P.Ong, T.V.Ramakrishnan, J.M.Tarascon and K.Remschnig, Phys.Rev.Lett. \textbf{67},761(1991);

\bibitem{22}P.A.Lee and T.V.Ramakrishnan, Rev.Nod.Phys. \textbf{57}, 287(1985);

\bibitem{23}S.I.Vedeneev and D.K.Maude, Phys.Rev.B \textbf{70}, 184524(2004);

\bibitem{24}F.Sonder,B.Chakoumakos, and B.Sales, Phys.Rev.B \textbf{40}, 6872(1989);
\end{thebibliography}
\end{document}